\documentclass[aps,prb,10pt,reprint,groupedaddress,superscriptaddress]{revtex4-2}

\usepackage[usenames,dvipsnames]{xcolor}
\usepackage{amssymb}
\usepackage{graphicx}
\usepackage{amsmath}
\usepackage{txfonts}
\usepackage[bookmarks=true,colorlinks,citecolor=blue,urlcolor=blue]{hyperref}
\usepackage{braket}
\usepackage{babel}
\usepackage{blindtext}
\usepackage[normalem]{ulem}

\usepackage{bm}  
	
\newcommand{\rom}[1]{\uppercase\expandafter{\romannumeral #1\relax}}

\newcommand{\jjpmodel}{J$\text{-}$J'}

\newcommand{\hpmodel}{[\frac{\pi}{2}, 0]}
\newcommand{\zpmodel}{[0, \pi]}
\newcommand{\zzmodel}{[0, 0]}

\newcommand{\orderphas}{$\sqrt{3}\times\sqrt{3}$ }

\newcommand{\PsiIstar}{$|{\Psi_{\text{I}}}^{*}\rangle$}
\newcommand{\PsiS}{$|\Psi_{\text{S}}\rangle$}
\newcommand{\PsiSstar}{$|{\Psi_{\text{S}}}^{*}\rangle$}

\begin{document}

\title{Possible chiral spin liquid state in the $S=1/2$ kagome Heisenberg model}

\newcommand{\RCCS}{\affiliation{Computational Materials Science Research Team, RIKEN Center for Computational Science (R-CCS), Kobe, Hyogo, 650-0047, Japan}}
\newcommand{\RQC}{\affiliation{Quantum Computational Science Research Team, RIKEN Center for Quantum Computing (RQC), Wako, Saitama, 351-0198, Japan}}
\newcommand{\KITS}{\affiliation{Kavli Institute for Theoretical Sciences, University of Chinese Academy of Sciences, Beijing 100190, China}}
\newcommand{\TUM}{\affiliation{Technical University of Munich, TUM School of Natural Sciences, Physics Department, 85748 Garching, Germany}}
\newcommand{\TUD}{\affiliation{Institut f\"ur Theoretische Physik, Technische Universit\"at Dresden, 01062 Dresden, Germany}}
\newcommand{\IOP}{\affiliation{Institute of Physics, Chinese Academy of Sciences, Beijing 100190, China}}

\author{Rong-Yang Sun} \thanks{These two authors contributed equally.} \RCCS  \RQC \KITS 
\author{Hui-Ke Jin}  \thanks{These two authors contributed equally.} \TUM

\author{Hong-Hao Tu} \email{hong-hao.tu@tu-dresden.de} \TUD

\author{Yi Zhou} \email{yizhou@iphy.ac.cn} \IOP \KITS 

\date{\today}
	
\begin{abstract}
	The nature of the ground state for the $S = 1/2$ Kagome Heisenberg antiferromagnet (KHAF) has been elusive. We revisit this challenging problem and provide numerical evidence that its ground state is a chiral spin liquid. Combining the density matrix renormalization group method with analytical analyses, we demonstrate that the previously observed chiral spin liquid phase in the KHAF with longer-range couplings is stable in a broader region of the phase diagram. We characterize the nature of the ground state by computing energy derivatives, revealing the ground-state degeneracy arising from the spontaneous breaking of the time-reversal symmetry, and targeting the semion sector. We further investigate the phase diagram in the vicinity of the KHAF and observe a $\sqrt{3}\times\sqrt{3}$ magnetically ordered phase and two valence-bond crystal phases.	
\end{abstract}

\maketitle

\noindent \textbf{INTRODUCTION}

\noindent The search for quantum spin liquids (QSLs) has been an active endeavor in condensed matter physics for decades~\cite{Anderson73,Anderson87,Lee08,Balents10,Savary2016,QSLRMP,Knolle2019,Broholm2020}. 
In contrast to magnetically ordered states, which are characterized by spontaneous symmetry breaking and local order parameters, this novel state of matter manifests itself in other exotic ways, such as long-range quantum entanglement and fractional spin excitations, which are represented by emergent particles (spinons) and gauge fields. In particular, a gapped QSL must be associated with a non-trivial topological order~\cite{Wen91,Hastings2004}.
A chiral spin liquid (CSL) is defined as a topological QSL that breaks time-reversal symmetry (TRS) and parity symmetry. It was first proposed by Kalmeyer and Laughlin~\cite{Kalmeyer87,Kalmeyer89} that the bosonic fractional quantum Hall state with filling factor $\nu=1/2$ could be the ground state of some frustrated Heisenberg antiferromagnets in two dimensions. Soon after that, Wen, Wilczek, and Zee~\cite{Wen89} introduced the spin chirality $E_{123}\equiv{}\mathbf{S}_1\cdot\left(\mathbf{S}_2\times\mathbf{S}_3\right)$ to characterize such a TRS-breaking QSL state, which they named CSL.

With strong geometric frustration and concrete quantum material realizations, the $S=1/2$ Kagome Heisenberg antiferromagnet (KHAF) has been regarded as one of the promising candidates to host QSLs~\cite{Mendel2007,Helton2007,lee2007,han2012,Norman16}. Theoretically, it has been generally accepted that the ground state of the KHAF with nearest-neighbor (NN) interactions is a QSL state~\cite{QSLRMP,Norman16}, although a valence-bond crystal (VBC) with a 36-site unit cell has also been proposed~\cite{Marston91,Singh2007,Nikolic2003,Vidal10}.
However, despite extensive research~\cite{Sachdev1992,Leung93,Lecheminant97,Mila98,Waldtmann98,Sindzingre09,Lauchli11,Lauchli19,Wietek2020,Jiang2008,yan2011,Depenbrock2012,jiang2012,nishimoto2013,Mei2017,Ran2007,Iqbal2013,Iqbal2014,Iqbal2015,YCHe2017,Liao2017,Jahromi2020,Gotze2011,TLi2018,Thoenniss2020}, the nature of the QSL state is still a mystery. In particular, the issue of ``to gap or not to gap" has not yet been settled down unambiguously.
While exact diagonalization~\cite{Leung93,Lecheminant97,Mila98,Waldtmann98,Sindzingre09,Lauchli11,Lauchli19}, density matrix renormalization group (DMRG)~\cite{Jiang2008,yan2011,Depenbrock2012,jiang2012,nishimoto2013}, and a tensor network variational calculation~\cite{Mei2017} indicate a finite energy gap, variational Monte Carlo~\cite{Ran2007,Iqbal2013,Iqbal2014,Iqbal2015}, infinite DMRG~\cite{YCHe2017}, and further tensor network studies~\cite{Liao2017,Jahromi2020} favor a gapless ground state, i.e., the U(1) Dirac QSL~\cite{Ran2007}.

Meanwhile, some early mean-field theory calculations proposed the possibility of a CSL~\cite{Marston91,Yang1993,Messio2012,Messio2013}. Further studies have shown that a CSL can be stabilized in the KHAF with chiral three-spin interactions~\cite{Bauer2014}, or with 2nd and 3rd NN Ising~\cite{YCHe2014Prl} or Heisenberg couplings~\cite{gong2014emergent,gong2015global,Wietek2015,Hu2015}, where the nature of this CSL was found to be of the $\nu = 1/2$ Kalmeyer-Laughlin type.

\begin{figure}[tb]
	\centering
	\includegraphics[width=1.0\linewidth] {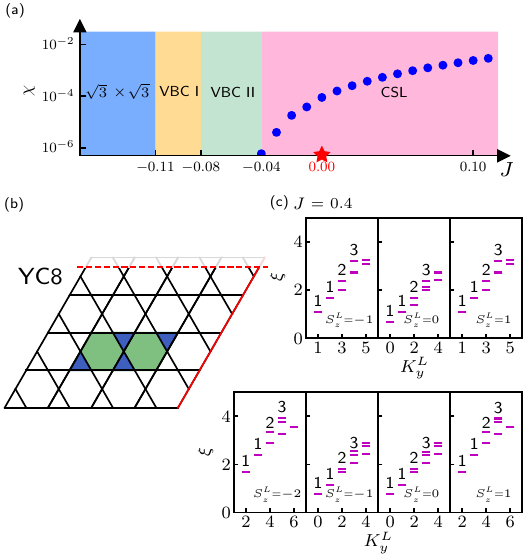}
    \caption{\label{fig:fig1} (a) The phase diagram of the $\jjpmodel$ model consisting of a $\sqrt{3}\times\sqrt{3}$ ordered phase, two VBC phases [VBC~I and VBC~II with different patterns, see Figs.~\ref{fig:fig3}(e) and (f)], and a CSL phase with finite spin chirality $\chi$. Note that the KHAF at $J' = 0$ (red star) also lies in the CSL phase with $\chi\approx{}2\times{}10^{-4}$. (b) YC8 cylinder for the Kagome lattice, where the dashed red line represents periodic boundary condition, and the flat right edge is highlighted by the solid red line. The colored area denotes a unit cell of the $\hpmodel$ mean-field ansatz, with $\pi/2$ (zero) flux through the elementary triangles (hexagons).  (c) Entanglement spectra of the left half of the cylinder, labeled by $K^{L}_{y}$ (momentum along the $y$-direction) and $S^{L}_{z}$ (total $z$-component spin), for ground states in the (upper panel) identity and (lower panel) semion sectors on the 152-site YC8 cylinder with $J'=0.4$.}
\end{figure}

In this work, we reexamine the evolution of the ground state of the KHAF with equal 2-nd and 3-rd NN Heisenberg couplings. We use the DMRG method~\cite{white1992,white1993,AnnalPhysSchollwock} (up to bond dimension $D=18000$) and analytical analyses to map out its ground-state phase diagram [see Fig.~\ref{fig:fig1}(a)], which consists of a $\sqrt{3}\times\sqrt{3}$ ordered phase, two distinct VBC phases, and a CSL phase. The most striking observation is that the ground state of the KHAF with only NN interactions lies in the aforementioned CSL phase, which is supported by the calculations of energy susceptibility, wave function fidelity, spin chirality, ground-state degeneracy, and the existence of topologically degenerate ground states in the semion sector. It is further shown that the CSL phase extends into the region where the 2nd and 3rd NN couplings are ferromagnetic.\\

\noindent \textbf{RESULT}

\noindent \textbf{Model}

\noindent We consider the KHAF with SU(2)-symmetric longer-range couplings (referred to as the $\jjpmodel$ model hereafter)
\begin{equation}
	\label{eq:H1}
	H = J\sum_{\langle ij \rangle_{1}} \mathbf{S}_{i}\cdot\mathbf{S}_{j}+J'\left(\sum_{\langle ij \rangle_{2}} \mathbf{S}_{i}\cdot\mathbf{S}_{j} + \sum_{\langle ij \rangle_{3}} \mathbf{S}_{i}\cdot\mathbf{S}_{j}\right)~,
\end{equation}
where $\langle ij \rangle_{\gamma}$ $(\gamma=1,2,3)$ denote the $\gamma$-th NN bonds on the Kagome lattice (the 3rd NN bonds are defined within hexagons only).
While it is also interesting to explore the general phase diagram when the 2-nd and 3-rd NN bond couplings are not equal~\cite{YCHe2014Prl,gong2014emergent,gong2015global},  we restrict ourselves to the model in Eq.~\eqref{eq:H1}, which, as we shall see below, proves sufficient for understanding the interpolation to the KHAF at $J'=0$. Henceforth we set $J = 1$ as the energy unit and focus on the regime of $-0.15 \leq J' \leq 0.4$ in the YC cylinder geometry [see Fig.~\ref{fig:fig1}(b)].\\

\noindent \textbf{Chiral spin liquid phase}

\noindent 
Let us start the discussion of the Hamiltonian in Eq.~\eqref{eq:H1} from its relatively well-understood CSL phase with moderately large $J'\gtrsim 0.08$~\cite{gong2014emergent,gong2015global}, for which much insight can be gained from a Gutzwiller projected wave function~\cite{Hu2015,Wietek2015}. This wave function is constructed in a fermionic parton representation for $S=1/2$ spins, 
$S^{a}_i=\frac{1}{2}\bm{c}_i^\dag\sigma^{a}\bm{c}_i$ $(a=x,y,z)$, where $\sigma^a$ are Pauli matrices and $\bm{c}_i=(c_{i,\uparrow},c_{i,\downarrow})^T$ are fermion annihilation operators. 
The physical Hilbert space of $S=1/2$ is restored by imposing a local single-occupancy constraint $\sum_{\sigma=\uparrow,\downarrow}c^\dag_{i{}\sigma}c_{i{}\sigma}=1$.

In the parton representation, a mean-field Hamiltonian, parameterized by link variables $\chi_{ij}=e^{\phi_{ij}}$ on the 1st NN bonds, takes the form of~\cite{Ran2007}
\begin{equation}
\label{eq:pt_mf}
	H_{\text{MF}} = \sum_{\langle{}ij\rangle_1}\sum_{\sigma=\uparrow,\downarrow}(\chi_{ij}c^{\dagger}_{i\sigma}c_{j\sigma} + \text{H.c.})~.
\end{equation}
Due to the SU(2) gauge redundancy~\cite{PSG0,PSG1,PSG2,PSG3}, Gutzwiller projected states obtained from Eq.~\eqref{eq:pt_mf} are distinguished by the gauge fluxes threading through elementary triangles and hexagons on the Kagome lattice [see Fig.~\ref{fig:fig1}(b)]. A CSL state is of particular interest that corresponds to the ansatz with $\frac{\pi}{2}$-flux through triangles and zero-flux through hexagons, dubbed as $\hpmodel$ state. Note that other parton \protect{Ans\"atze} in this work will be labeled in the same scheme. 

With a cylindrical boundary condition, the $\hpmodel$ state exhibits four exact boundary zero modes, $d_{L\sigma}^{\dagger}$ and $d_{R\sigma}^{\dagger}$, where $L$ ($R$) denotes the left (right) boundary of the cylinder. For the CSL state, the minimally entangled states (anyon eigenbasis)~\cite{zhang2012quasiparticle} are constructed as follows~\cite{tu2013b,MPOMPS1}:
\begin{equation}
    \label{eq:zeromodes}
    |\Psi_{1}\rangle = \hat{P}_{G} d_{L\uparrow}^{\dagger}d_{L\downarrow}^{\dagger}|\Phi\rangle, \quad |\Psi_{2}\rangle = \hat{P}_{G} d_{L\uparrow}^{\dagger}d_{R\downarrow}^{\dagger} |\Phi\rangle~,
\end{equation}
where $|\Phi\rangle$ is a Fermi sea state with all negative energy modes of Eq.~\eqref{eq:pt_mf} being occupied, and $\hat{P}_{G}$ is the Gutzwiller projection operator imposing the single-occupancy constraint.
With the MPO-MPS method~\cite{MPOMPS1,Jin2021}, the CSL anyon eigenbasis in Eq.~\eqref{eq:zeromodes} can be converted into matrix product states (MPSs) with high fidelity.

For the $\jjpmodel$ model in the CSL phase, two topologically degenerate ground states, denoted by $|\Psi_{\text{I}}\rangle$ ($|\Psi_{\text{S}}\rangle$) in the identity (semion) sector, can be obtained by initializing DMRG with their parton counterpart $|\Psi_{1}\rangle$ ($|\Psi_{2}\rangle$)~\cite{Jin2021BDMRG,JYChen2021}. For a YC8 cylinder with 152 sites,
the total ground-state energy difference (including boundary contributions)
between two topological sectors is $\Delta E_0 \approx 0.33\ (0.27)$ for $J' = 0.4\ (0.2)$, and the two ground states are orthogonal to each other, as indicated by their overlap $|\langle \Psi_{\text{I}} | \Psi_{\text{S}}\rangle| \sim$ $10^{-11}\ (10^{-9})$.
We note that the time-reversal partner of $|\Psi_{\text{I}}\rangle$ ($|\Psi_{\text{S}}\rangle$), denoted by \PsiIstar\ (\PsiSstar), has the same energy as $|\Psi_{\text{I}}\rangle$ ($|\Psi_{\text{S}}\rangle$) due to the spontaneous TRS breaking.
As shown in
Fig.~\ref{fig:fig1}(c)
, the entanglement spectrum~\cite{Haldane2008,Qi2012} of $|\Psi_{\text{I}}\rangle$ ($|\Psi_{\text{S}}\rangle$) displays the characteristic counting $\{1,3,4,7,...\}$ ($\{2,2,6,8,...\}$), which suggests the chiral SU(2)$_{1}$ Wess-Zumino-Witten model being the edge conformal field theory. These results firmly validate the existence of a CSL phase of Kalmeyer-Laughlin type in this model.\\

\noindent \textbf{Spontaneous TRS breaking at $J'=0$}

\noindent
After examining the CSL phase at relatively large $J'$, we now show that this CSL phase is adiabatically connected to the ground state at $J'=0$ point. 
For this purpose, we carry out DMRG calculations that are initialized with randomly-generated MPSs (dubbed as Random-DMRG) as well as specific wave functions (dubbed as Boosted-DMRG~\cite{Jin2021BDMRG}). For Boosted-DMRG, we adopt either the parton ansatz~\eqref{eq:zeromodes} or post-optimized MPSs of the neighborhood Hamiltonian in parameter space (see also a related adiabatic DMRG approach~\cite{gong2014emergent,sun2021metal}). Nevertheless, the ground-state energies obtained by Random- and Boosted-DMRG are almost identical (see Supplementary Note 1), both of which agree with the results in Ref.~\cite{gong2015global}. For the KHAF with $J'=0$, the ground-state energy per site is found to be $-0.4383(6)$, which also agrees with the best available DMRG results~\cite{yan2011,Depenbrock2012}.

\begin{figure}[tb]
	\centering
	\includegraphics[width=1.0\linewidth]{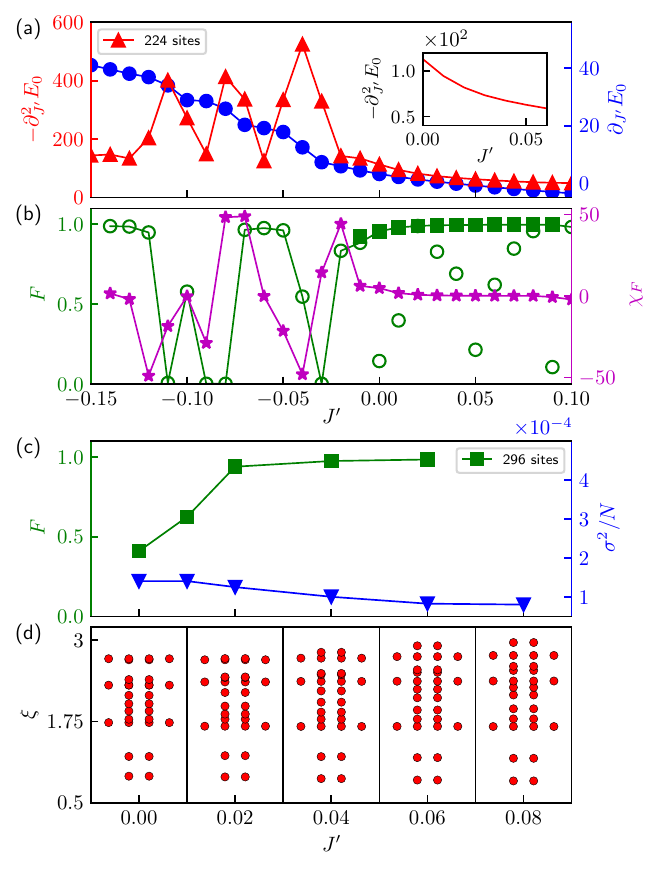}
	\caption{\label{fig:fig2}
		 (a) The first- (blue dots) and second-order (red triangles) derivatives of the ground-state energy $E_0$ versus $J'$ on the 224-site YC8 cylinders. The inset shows a zoom-in of the second-order energy derivatives around the $J'=0$ point.
		(b) The neighboring wave function overlap $\langle\Psi(J')|\Psi(J'-\Delta{}J')\rangle$ ($\Delta{}J'=0.01$) and the corresponding susceptibility versus $J'$. The empty circles are obtained by Random-DMRG, and the full squares are obtained by Boosted-DMRG. Two states at the same $J'$ have quasi-degenerated energies (see Supplementary Note 1).
		(c) Neighboring wave function overlaps (green squares) and energy variances (blue triangles) of semion-sector ground states in 296-site YC8 cylinders obtained by Boosted-DMRG. (d) The low-lying entanglement spectra of section-sector ground states remain qualitatively similar as $J'$ decreases.}
\end{figure}

\begin{figure}
    \centering
    \includegraphics[width=1.0\linewidth]{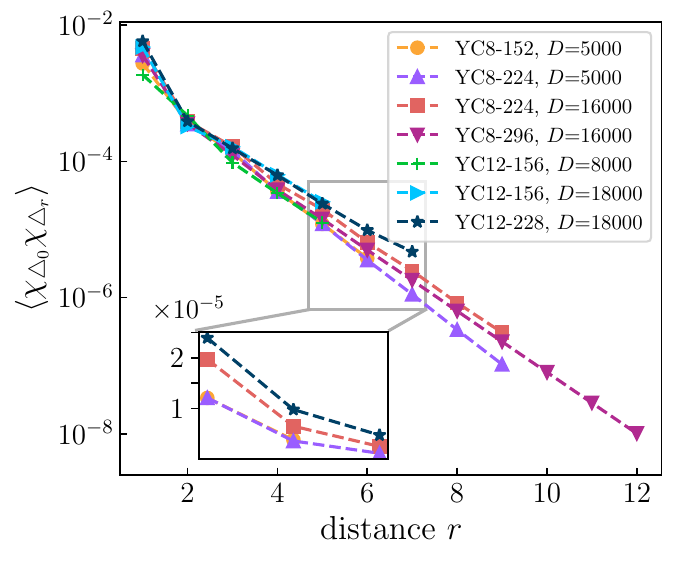}
    \caption{\label{fig:cc}
        Semi-Log plot of chiral-chiral correlation function as a function of distance $r$ for the $J' = 0$ model on YC8 and YC12 cylinders. Inset: Zoom-in of the area bounded by gray box with a linear plot.}
\end{figure}

For $-0.15 \leq J' \leq 0.4$, the first-order derivative of the ground-state energy $E_0$ does not show any singularities, as shown in Fig.~\ref{fig:fig2}(a). However, we observe that the second-order derivative of $E_0$ exhibits three sharp peaks [see Fig.~\ref{fig:fig2}(a)]. The location of these peaks indicates that there is no (at least neither first nor second-order) phase transition within the region of $0\leq{}J'\leq{}0.1$. We emphasize that these results are stable against the system-size and bond-dimension scaling (see Supplementary Note 6) and indeed, are consistent with a previous DMRG result~\cite{gong2015global}. In contrast, previous DMRG studies~\cite{gong2014emergent,gong2015global} suggest a critical point at $J'_c\approx0.08$ that is identified by vanishing spin chirality. This contradiction can be resolved as follows~(see Supplementary Note 2):
Consider the spin chirality $\chi_{ijk}(\geq{}0)$ defined on an elementary triangle $\triangle_{ijk}$ [see Fig.~\ref{fig:fig1}(b)], we find that (i) $\chi_{ijk}$ is sizable for a state deep inside the CSL phase, e.g., $\chi_{ijk}\sim{}0.06$ at $J'=0.2$; (ii) $\chi_{ijk}$ will decrease rapidly as $J'$ decreases and exceed $J'=0.1$, e.g., $\chi_{ijk}\sim{}6\times{}10^{-4}~ (2\times{}10^{-4})$ at $J'=0.08~(0)$, which is small but does not vanish numerically (note that the truncation error in our DMRG calculations is $\lesssim{}10^{-6}$), and is stable against finite-size scaling (see Supplementary Note 2); (iii) $\chi_{ijk}$ continues to decrease to a numerical zero, $\chi_{ijk}\sim{}6\times{}10^{-7}$ at $J'=-0.04$, around which CSL to VBC II phase transition occurs [see Figs.~\ref{fig:fig1}(a) and \ref{fig:fig2}].
Moreover, in Fig.~\ref{fig:cc}, we show the real-space chiral-chiral correlation function at $J'=0$. 
Comparing $J'=0$ and $J'=0.2$ (see the results of $J'=0.2$ in Supplementary Note 2 and also in Ref.~\cite{gong2015global}) data with various parameters, one can conclude that (i) the long-ranged correlation of spin chirality becomes evident only 
when the circumference and bond dimensions are sufficiently large. This phenomenon holds true even when the system is deep inside the CSL phase, namely, $J'=0.2$; (ii) Increasing the circumference of cylinders and/or bond dimensions can enhance the chiral-chiral correlation obtained by DMRG, thereby amplifying the strength of spin chirality. These observations strongly suggest that even at $J' = 0$, the chiral-chiral correlation tends to be long-ranged on cylinders with sufficiently large circumferences, provided that the bond dimensions are adequately sized to catch this feature. 

In addition to the energetics, we have also calculated the wave function fidelity $F(J') = |\langle \Psi(J') | \Psi(J'-\Delta{}J') \rangle |$ and the corresponding susceptibility $\chi_{F}\equiv{}\text{d}F/\text{d}J'$, where $|\Psi(J') \rangle$ is the ground state obtained by Random- or Boosted-DMRG initialized with $|\Psi(J'+\Delta{}J')\rangle$. Here we adopt $\Delta{}J'=0.01$. For $J'\leq{}-0.04$, the kinks in $F$ and singularities in $\chi_{F}$, as shown in Fig.~\ref{fig:fig2}(b), are in agreement with the location of the sharp peaks in the second-order derivative of $E_0$.

For $-0.03 < J' \leq 0.09$, the fidelity $F$ obtained from Random-DMRG (using \textit{real} MPSs) fluctuates strongly without any regular pattern. This seemingly odd result indicates that the system either has degenerate ground states (from which DMRG chooses their superposition, which varies randomly at different $J'$) or hosts a pile of phase transitions within this narrow region. As we mentioned earlier, $|\Psi_{\text{I}}\rangle$ and its time-reversal partner $|{\Psi_{\text{I}}}^{*}\rangle$ are Kramers degenerate ground states in the CSL phase, whose \textit{real} combinations are $|\Phi_{\text{I}}\rangle = (|\Psi_{\text{I}}\rangle + |{\Psi_{\text{I}}}^{*}\rangle)/\sqrt{2}$ and $|\tilde{\Phi}_{\text{I}}\rangle = i(|\Psi_{\text{I}}\rangle - |{\Psi_{\text{I}}}^{*}\rangle)/\sqrt{2}$. As usual, Random-DMRG using real MPSs would find their superposition $\cos(\theta) |\Phi_{\text{I}}\rangle + \sin(\theta) |\tilde{\Phi}_{\text{I}}\rangle$. Though DMRG is typically biased towards a low-entanglement combination, here the state is an equal weight superposition of two minimally entangled states, i.e., \protect{$(e^{i\theta}|\Psi_{\text{I}}\rangle + e^{-i\theta} |{\Psi_{\text{I}}}^{*}\rangle)/\sqrt{2}$}. Therefore, DMRG has no preference over a particular $\theta$ which is not under control and varies with $J'$. This would lead to a fluctuating fidelity $F$ in Fig.~\ref{fig:fig2}(b). Furthermore, we note that $\theta$ can be fixed (or slowly varying) by performing a series of Boosted-DMRG calculations, i.e., the ground-state search for coupling $J'$ is carried out by initializing DMRG with the post-optimized MPS for coupling $J'+\Delta{}J'$. This adiabatic procedure can be done recursively from $J' = 0.1$ to $J'=-0.03$ and, indeed, the fidelity $F$ now becomes a plateau close to $1$, as shown in Fig.~\ref{fig:fig2}(b). This is sharp evidence for the existence of degenerate ground states and also rules out the multiple-transition scenario.

Generally, the two MPSs obtained from Random-DMRG and Boosted-DMRG are not orthogonal to each other and can be used to determine the two-dimensional ground-state subspace spanned by $|\Phi_{\text{I}}\rangle$ and $|\tilde{\Phi}_{\text{I}}\rangle$. This can be formulated as a variational problem by diagonalizing the Hamiltonian in this subspace, which yields two orthogonal states and their energies. For $-0.03 \leq J' \leq 0.08$, the relative difference between these two energies is smaller than $3\times 10 ^{-5}$ (see Supplementary Note 1). Also, we compute the energy variance $\sigma^{2}\equiv{}\langle{}H^2\rangle-\langle{}H\rangle^2$ with respect to DMRG-optimized ground states and find that $\sigma^{2} / N$ is smaller than $2\times 10 ^{-4}$. Combining these observations with quasi-degenerate energy (
see Supplementary Note 1), we conclude that the states obtained by Random- and Boosted-DMRG indeed form a two-dimensional ground-state space. It is also worth mentioning that the energy difference between these two states is smaller than the gap estimates in previous studies \cite{yan2011,Lauchli19}.\\

\noindent \textbf{Topologically degenerate semion sector}

\noindent
Another essential signal of CSL around $J'=0$ is that \PsiS, the ground state in the semion sector on a 296-site YC8 cylinder, can be continuously evolved from $J'=0.2$ to $J'=0$ by using Boosted-DMRG. As shown in Fig.~\ref{fig:fig2}(c), the wave function fidelity $F$ between two neighboring-$J'$ \PsiS's is always close to unity for $0.02\leq{}J'\leq{}0.2$ and then drops to a smaller but finite value ($F\gtrsim{}0.4$). Note that this sudden drop is caused by a slight shift of edge semions in real space and does not correspond to a phase transition, as will be discussed in next paragraph.
Furthermore, the entanglement spectra of \PsiS's also show qualitative consistency in the region $0\leq{}J'\leq{}0.2$, as demonstrated in Fig.~\ref{fig:fig2}(d). Moreover, the tiny per-site energy variances of \PsiS's, $\sigma^{2}/N\sim{}10^{-4}$, suggests that \PsiS\ is an eigenstate of Hamiltonian~\eqref{eq:H1}. A study of the scaling behavior of $\sigma^2$ further confirms the robustness of our results (for details, see Supplementary Note 6).

We would like to delve into the sudden drop of the wave function fidelity starting at $J'\approx{}0.01$. It turns out that when $J'$ is closer and closer to $J'=0$ (or be precise, $J'=J_c\approx-0.04$), the finite-size energy gap between identity and semion sectors becomes smaller and smaller. 
Meanwhile, the edge semions at each boundary are more and more delocalized (see Supplementary Note 3), which increases the likelihood of these semions ``colliding'' with each other on finite cylinders. Thereby, it is more difficult to stabilize \PsiS\ as a ``metastable'' higher-energy state in DMRG, and the collapse of \PsiS\ into the identity-sector ground state $|\Psi_{\text{I}}\rangle$ is more likely to occur due to the DMRG energy optimization. This agrees with our observation that the evolution of \PsiS\ can be unstable when the length of the cylinder is not long enough. In fact, the evolution breaks down at around $J' < 0.02$ when examining a shorter 152-site YC8 cylinder (see Supplementary Note 3). 


\begin{figure*}[tb]
	\centering
	\includegraphics[width=\linewidth]{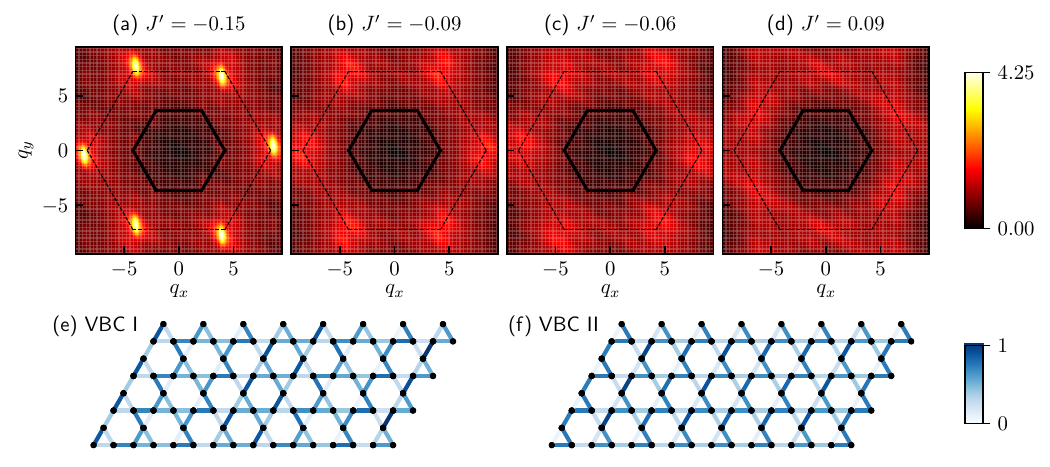}
	\caption{\label{fig:fig3}
	Top: The spin structure factor $S(\mathbf{q})$ on 224-site YC8 cylinders for (a) the \orderphas phase, (b) the VBC~I phase, (c) the VBC~II phase, and (d) the CSL phase. The solid (dashed) hexagon is the first (extended) Brillouin zone of the Kagome lattice.
	Bottom: Normalized absolute value of the NN spin-spin correlation $\langle \mathbf{S}_{i}\cdot\mathbf{S}_{j}\rangle$ on a 224-site YC8 cylinder for (e) the VBC~I phase and (f) the VBC~II phase. Note that the VBC~II phase is similar with the VBC phase in \cite{gong2015global}. The cylinders are shown with only central columns.}
\end{figure*}

For the most contentious KHAF with $J' = 0$, we have also considered two more Gutzwiller \protect{Ans\"atze} associated with Eq.~\eqref{eq:pt_mf}: (i) $\zpmodel$ state [a gapless U(1) Dirac QSL] and (ii) $\zzmodel$ state (a QSL with spinon Fermi surface)~\cite{Ran2007}. After converting them into MPSs, we find that their overlaps with the DMRG-optimized ground states are almost zero. When using these two \protect{Ans\"atze} for initializing DMRG, neither of them can improve the performance of DMRG (see Supplementary Note 4). Although this does not rule out the U(1) Dirac and spinon Fermi surface QSL scenarios, it gives a hint that neither $\zpmodel$ nor $\zzmodel$ ansatz is a good description for the ground state of the KHAF.\\

\noindent \textbf{Phase diagram}

\noindent
To investigate the nature of the other three phases for $J' \lesssim -0.04$, we calculate the NN spin-spin correlation $\langle \mathbf{S}_{i}\cdot\mathbf{S}_{j}\rangle$ and the spin structure factor
\begin{equation}
	\label{eq:sq}
	S(\mathbf{{q}})=\frac{1}{N} \sum_{i j}\langle\mathbf{S}_{i} \cdot \mathbf{S}_{j}\rangle e^{i \mathbf{q} \cdot (\mathbf{r}_{i}-\mathbf{r}_{j})}~,
\end{equation}
which are illustrated in Fig.~\ref{fig:fig3}.
For $J'<-0.11$, $S(\mathbf{{q}})$ exhibits clear peaks at the $K$ points in the extended Brillouin zone [see Fig.~\ref{fig:fig3}(a)], and the NN correlation $\langle \mathbf{S}_{i}\cdot\mathbf{S}_{j}\rangle$ is spatially uniform (not shown), which can be identified as the $\sqrt{3}\times\sqrt{3}$ spin-ordered phase~\cite{Sachdev1992,Kolley2015,Wietek2020}.
In the two intermediate phases ($-0.11 \leq J' \leq -0.04$), $S(\mathbf{{q}})$ are featureless [see Figs.~\ref{fig:fig3}(b) and (c)], indicating the absence of long-range magnetic order, and the NN correlation $\langle \mathbf{S}_{i}\cdot\mathbf{S}_{j}\rangle$ exhibits clear translational symmetry breaking patterns [see Figs.~\ref{fig:fig3}(e) and (f)], which are two different VBC phases (see more details in Supplementary Note 5) and dubbed as VBC~I and VBC~II in Fig.~\ref{fig:fig1}(a). Finally, in the CSL phase, $S(\mathbf{{q}})$ is featureless [see Fig.~\ref{fig:fig3}(d)] and the NN correlation $\langle \mathbf{S}_{i}\cdot\mathbf{S}_{j}\rangle$ is also spatially uniform (not shown), which are consistent with a QSL state. \\

\noindent \textbf{DISCUSSION}

\noindent
To summarize, we revisit the KHAF by combining the DMRG method and analytical analyses and provide clear evidence that the CSL phase in the KHAF with longer-range interactions extends to a broader region in the phase diagram and, most strikingly, includes the KHAF with only NN interactions. In the vicinity of the CSL phase, two distinct VBC phases and a $\sqrt{3}\times\sqrt{3}$ magnetically ordered phase emerge in the phase diagram (in order of decreasing $J'$).


Moreover, Gutzwiller projected wave function provides us a good initial ansatz to systematically prepare and study ground states in distinct topological sectors, without involving other technique tricks, such as pinning fields and flux insertion~\cite{YCHe2014Prb,gong2014emergent}.
Thanks to this great advantage, the ground state in the semion sector can be properly prepared, and then adiabatically evolved by DMRG all the way up to $J^{\prime}=0$.
Since the semion sector is a fingerprint of a Kalmeyer-Laughlin-type CSL, we argue that a gapped CSL appears to be the most promising candidate for the ground state at $J^\prime=0$.

Our work may also attract new future attention to this pendent issue. It would be desirable to observe the degenerate ground states at $J' = 0$ by, e.g., larger-scale DMRG calculations on wider cylinders or tensor network calculations in the thermodynamic limit. From the experimental side, the signature of semion excitations could be explored in real materials, even when the Kagome compounds, having interactions beyond the KHAF with NN interactions, may no longer have a CSL as their ground states.\\


\noindent \textbf{DATA AVAILABILITY}

\noindent
All data needed to evaluate the conclusions in the paper are present in the paper and/or the Supplementary Materials. The raw data sets used for the presented analysis within the current study are available from the corresponding authors on reasonable request.\\

\noindent {\bf ACKNOWLEDGEMENTS}

\noindent {\bf Acknowledgement} 

\noindent
We thank Shou-Shu Gong, Ying-Hai Wu, Shuo Yang, Hong Yao, and Zheng Zhu for the valuable discussions. R.-Y.S. is supported by the COE research grant in computational science from Hyogo Prefecture and Kobe City through Foundation for Computational Science. H.-K.J. is funded by the European Research Council (ERC) under the European Unions Horizon 2020 research and innovation program (grant agreement No. 771537). H.-H.T. is supported by the Deutsche Forschungsgemeinschaft (DFG) through project A06 of SFB 1143 (project No. 247310070). Y.Z. is supported by National Key Research and Development Program of China (No. 2022YFA1403403) and
National Natural Science Foundation of China (No.12274441, 12034004).
The numerical simulations in this work are based on the GraceQ project \cite{graceq}.\\ 

\noindent {\bf Author contributions}  

All authors contributed to conception, execution, and write-up of this project.\\

\noindent {\bf Competing interests} 

The authors declare that they have no competing interests.

\bibliography{m2kafm.bib}

\begin{widetext}
\begin{center}
\bf{ Supplemental Materials}
\end{center}

\setcounter{table}{0}   
\setcounter{figure}{0}
\setcounter{equation}{0}
\renewcommand{\thetable}{S\arabic{table}}
\renewcommand{\thefigure}{S\arabic{figure}}
\renewcommand{\theequation}{S\arabic{equation}}

\renewcommand{\figurename}{Supplementary Fig.}
\renewcommand\refname{Supplementary Reference}

This Supplemental Material includes more numerical details of the DMRG calculations. 
In Supplementary Note 1, we show an energetic comparison between Random-DMRG and Boosted-DMRG.
In Supplementary Note 2, we perform a systematical study on the spin chirality near the point of $J' = 0$. In Supplementary Note 3, we provide more details on the characterizations of the ground state in the semion sector.
In Supplementary Note 4, we compare the performance of DMRG simulations at $J'=0$ with initial MPSs prepared from random states and several parton ansatz. 
In Supplementary Note 5, we show the bond-bond correlation functions to further characterize two VBC phases.
In Supplementary Note 6, we discuss the system-size and bond-dimension scalings of our DMRG calculations.

\subsection*{Supplementary Note 1: Energetic comparisons and re-diagonalization}
We introduce the relative energy difference between two close energies, $E_{1}$ and $E_{2}$, as
\begin{equation}
\label{eq:engs_relative_diff}
\delta (E_{1}, E_{2}) \equiv \left| \frac{E_{1}-E_{2}}{(E_{1} + E_{2}) / 2} \right|~,
\end{equation}
to quantify their closeness. As mentioned in the main text, we obtained two almost degenerate ground states, $|\Psi_{\rm RD}\rangle$ and $|\Psi_{\rm BD}\rangle$, by using Random-DMRG and Boosted-DMRG, respectively. In the CSL phase, these two states are generally not orthogonal to each other. Denoting the variational energy of $|\Psi_{\rm RD}\rangle$ ($|\Psi_{\rm BD}\rangle$) as $E_{\rm RD}$ ($E_{\rm BD}$), the relative energy difference $\delta(E_{\rm RD},E_{\rm BD})$ is extremely small, as listed in Table~\ref{tab:csl_engs_relative_diff}.

\begin{table}[!ht]
\centering
\renewcommand\arraystretch{1.2}
\setlength\tabcolsep{0.4cm}
\caption{\label{tab:csl_engs_relative_diff}
	Relative energy difference between Random-DMRG and Boosted-DMRG states, $\delta(E_{\rm RD}, E_{\rm BD})$, and between two re-orthogonalized states, $\delta(\tilde{E}_{\rm 1}, \tilde{E}_{\rm 2})$, in the CSL phase. The data are obtained on $224$-site YC8 cylinders.
}
\begin{tabular}{ccc}
	\hline
	\hline
	\multicolumn{1}{c}{$J'$} & \multicolumn{1}{c}{$\delta(E_{\text{RD}}, E_{\text{BD}})$} & \multicolumn{1}{c}{$\delta(\tilde{E}_{\text{1}}, \tilde{E}_{\text{2}})$} \\ \hline
	$0.08$                     & $1.81\times{}10^{-9}$                                                   & $3.13\times{}10^{-7}$                                                              \\
	$0.07$                     & $1.92\times{}10^{-7}$                                                   & $2.39\times{}10^{-6}$                                                            \\
	$0.06$                     & $1.50\times{}10^{-7}$                                                   & $2.44\times{}10^{-6}$                                                              \\
	$0.05$                     & $9.96\times{}10^{-8}$                                                   & $1.92\times{}10^{-6}$                                                              \\
	$0.04$                     & $1.35\times{}10^{-7}$                                                   & $4.73\times{}10^{-6}$                                                                \\
	$0.03$                     & $9.97\times{}10^{-9}$                                                   & $2.82\times{}10^{-5}$                                                              \\
	$0.02$                     & $1.72\times{}10^{-7}$                                                   & $4.11\times{}10^{-6}$                                                              \\
	$0.01$                     & $1.29\times{}10^{-7}$                                                   & $3.44\times{}10^{-6}$                                                              \\
	$0$                        & $6.45\times{}10^{-8}$                                                  & $1.13\times{}10^{-5}$                                                              \\
	$-0.01$                    & $4.42\times{}10^{-7}$                                                   & $9.79\times{}10^{-7}$                                                             \\
	$-0.02$                    & $1.61\times{}10^{-6}$                                                   & $2.63\times{}10^{-6}$                                                              \\
	$-0.03$                    & $1.49\times{}10^{-6}$                                                   & $4.90\times{}10^{-6}$                                                               \\
	\hline
	\hline
\end{tabular}
\end{table}

Meanwhile, with two non-orthogonal MPSs $|\Psi_{\text{RD}}\rangle$ and $|\Psi_{\text{BD}}\rangle$ in hand, we carry out a further variational optimization to obtain re-orthogonalized states and corresponding energies. This optimization is formulated as a generalized eigenvalue problem by constructing the following $2\times{}2$ matrices:
\begin{equation}
H_{2}=\left(\begin{array}{cc}
	\langle\Psi_{\text{RD}}|H|\Psi_{\text{RD}}\rangle & \langle\Psi_{\text{RD}}|H|\Psi_{\text{BD}}\rangle \\
	\langle\Psi_{\text{BD}}|H|\Psi_{\text{RD}}\rangle & \langle\Psi_{\text{BD}}|H|\Psi_{\text{BD}}\rangle
\end{array}\right),
\end{equation}
and
\begin{equation}
N_{2}=\left(\begin{array}{cc}
	\langle\Psi_{\text{RD}}|\Psi_{\text{RD}}\rangle & \langle\Psi_{\text{RD}}|\Psi_{\text{BD}}\rangle \\
	\langle\Psi_{\text{BD}}|\Psi_{\text{RD}}\rangle & \langle\Psi_{\text{BD}}|\Psi_{\text{BD}}\rangle
\end{array}\right).
\end{equation}
By solving the generalized eigenvalue problem, namely, $H_2{}x=E{}N_2{}x$, we can obtain two re-diagonalized eigen-energies, $\tilde{E}_{\text{1}}$ and  $\tilde{E}_{\text{2}}$. The relative energy difference $\delta(\tilde{E}_{\text{1}}, \tilde{E}_{\text{2}})$, with respect to newly obtained $\tilde{E}_{\text{1}}$ and  $\tilde{E}_{\text{2}}$, is also listed in Table~\ref{tab:csl_engs_relative_diff}. We can see that these energies are still very close to each other, implying that $|\Psi_{\text{RD}}\rangle$ and $|\Psi_{\text{BD}}\rangle$ are indeed the two states in the ground state manifold.

\subsection*{Supplementary Note 2: Spin chirality}

\begin{figure}
\centering
\includegraphics[width = 0.8\linewidth]{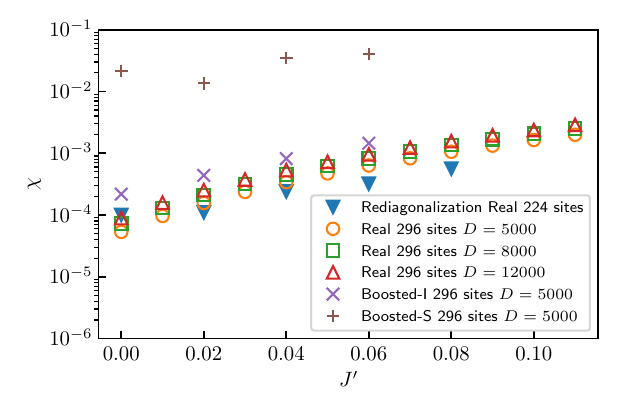}
\caption{\label{fig:c_scaling}
	The spin chirality $\chi$ calculated with different bond dimension $D$ and system size $N$. (i) The blue down-triangle is obtained by the re-diagonalization method. (ii) The orange circle, green square, and red triangle are evaluated from the chiral-chiral correlation function. (iii) The purple cross and brown plus are directly calculated by Boosted-DMRG (complex-number MPS). Here Boosted-I (Boosted-S) stands for the ground state in the identity (semion) sector.
}
\end{figure}

In order to characterize the spin chirality, we introduce the three-spin operator on an elementary triangle $\bigtriangleup_{ijk}$ [see the elementary triangles in Fig.~1(b) of the main text] as
\begin{equation}
E_{ijk} = \mathbf{S}_i\cdot\left(\mathbf{S}_j\times\mathbf{S}_k\right).
\end{equation}
A nonzero spin chirality in the ground state provides direct evidence of the spontaneous breaking of time-reversal symmetry, which is useful for characterizing the chiral spin liquid phase in the $J-J'$ model. Here, we calculate the spin chirality by the following three independent methods.

\textit{Method I: Re-diagonalize two degenerated states.---} Notice that $E_{ijk}$ is purely imaginary and Hermitian, indicating that $\langle\psi|E_{ijk}|\psi\rangle=0$ for any real state $|\psi\rangle$. In the CSL phase, we can obtain two orthogonal degenerate ground states, $|\tilde{\Phi}_{1}\rangle$ and $|\tilde{\Phi}_{2}\rangle$, by the re-diagonalization method introduced in Supplementary Note 1. Because both $|\tilde{\Phi}_{1}\rangle$ and $|\tilde{\Phi}_{2}\rangle$ are real wave functions, the expectation values of $E_{ijk}$ with respect to those two states must be zero. Nevertheless, we can evaluate the spin chirality by introducing the following $2\times{}2$ matrix:
\begin{equation}
\chi_2(ijk)=\left(\begin{array}{cc}
	0 & \langle\tilde{\Phi}_1|E_{ijk}|\tilde{\Phi}_2\rangle \\
	\langle\tilde{\Phi}_2|E_{ijk}|\tilde{\Phi}_1\rangle & 0 \end{array}\right).
\end{equation}
It is obvious that $\chi_2(ijk)$ is Hermitian and has symmetric eigenvalues with respect to zero, namely, $\pm{}\chi_{ijk}$ ($\chi_{ijk}>0$), corresponding to chiral and anti-chiral ground states, respectively. Hereafter we choose the elementary triangle $\bigtriangleup_{ijk}$ in the center of a YC8 cylinder to compute $\chi_{ijk}$. Notice that there are two types of elementary triangles corresponding to up- and down-triangles and they give rise to almost the same results.
To simplify the notation, we drop the subscripts and use $\chi$ to represent the spin chirality averaged by measuring typical elementary triangles.
For relatively large $J'$ ($J'=0.2$), we find that $\chi \approx{}0.06$ which is very close to the result obtained in Ref.~\cite{gong2015global}.
For the region near $J' = 0$, as shown in Supplementary Fig.~\ref{fig:c_scaling}, $\chi_{ijk}$ decreases as $J'$ decreases. At the point of $J'=0$, we find that $\chi\approx{}2.0\times{}10^{-4}$. As a sharp comparison, the spin chirality decreases to $\chi\approx{}6.0\times{}10^{-7}$ at $J' = -0.04$.

\textit{Method II: Extract spin chirality from chiral-chiral correlation functions.---} Although the scalar chirality order parameter is zero for real states, the spin chirality can be approximately detected by measuring the chiral-chiral correlation function as
\begin{equation}
\chi = \sqrt{|\langle E_{i_{0}j_{0}k_{0}} E_{i_{d}j_{d}k_{d}} \rangle|}~,
\end{equation}
where $i_{0}j_{0}k_{0}$ ($i_{d}j_{d}k_{d}$) indicates three sites in the starting (ending) elementary triangle. Therefore, $d$ indicates the distance between these two elementary triangles. In the thermodynamic limit, $\chi$ is finite for a time-reversal symmetry-breaking state even if $d$ is large. On the finite-length cylinders, we choose two elementary triangles with a sufficiently large distance. In order to suppress the boundary effect, we chose the case with the starting elementary triangle being at $N_{x} / 4$ and the ending elementary triangle $3N_{x} / 4$, where $N_x$ is the length of the cylinder. 
Note that the same method is also employed in Ref.~\cite{gong2015global} as the major approach to determine spin chirality. The corresponding data at $0 \leq J' \leq 0.11$ on a YC8 cylinder with $N=296$ and $N_{x} = 24$ is illustrated in Supplementary Fig.~\ref{fig:c_scaling}.

\textit{Method III: Directly measure spin chirality in complex-number MPS.---} In principle, the spin chirality $\chi$ does not necessarily vanished in a complex-number MPS. Therefore, the spin chirality can be determined by directly calculating $\chi_{ijk}=\langle\Psi_{\rm I(S)}|E_{ijk}|\Psi_{\rm I(S)}\rangle$, where $|\Psi_{\rm I(S)}\rangle$ is the ground state obtained by Boosted-DMRG initialized with a parton ansatz of CSL with $[\pi/2, 0]$ (see $|\Psi_{1(2)}\rangle$ in the main text). 
Because the Hamiltonian preserves the time-reversal symmetry, in finite DMRG calculations, the time-reversal partner, $|\Psi_{\rm I(S)}^*\rangle$, will be gradually mixed into the state by the numerical fluctuations introduced by the optimization process in DMRG. 
Therefore, the spin chirality measured using this method is smaller than the actual value of spin chirality for a CSL state, hence providing a lower bound. The data evaluated by method III are also illustrated in Supplementary Fig.~\ref{fig:c_scaling} for ground states in identity and semion sectors.

In Supplementary Fig.~\ref{fig:c_scaling}, the above three independent checks provide a consistent result. The spin chirality at $J'=0$ is around $10^{-4}$ and increases with the system size $N$ and the bond dimension $D$, suggesting that a small but finite spin chirality at $J'=0$ persists in the thermodynamic limit.


\begin{figure}
\centering
\includegraphics[width = 0.9\linewidth]{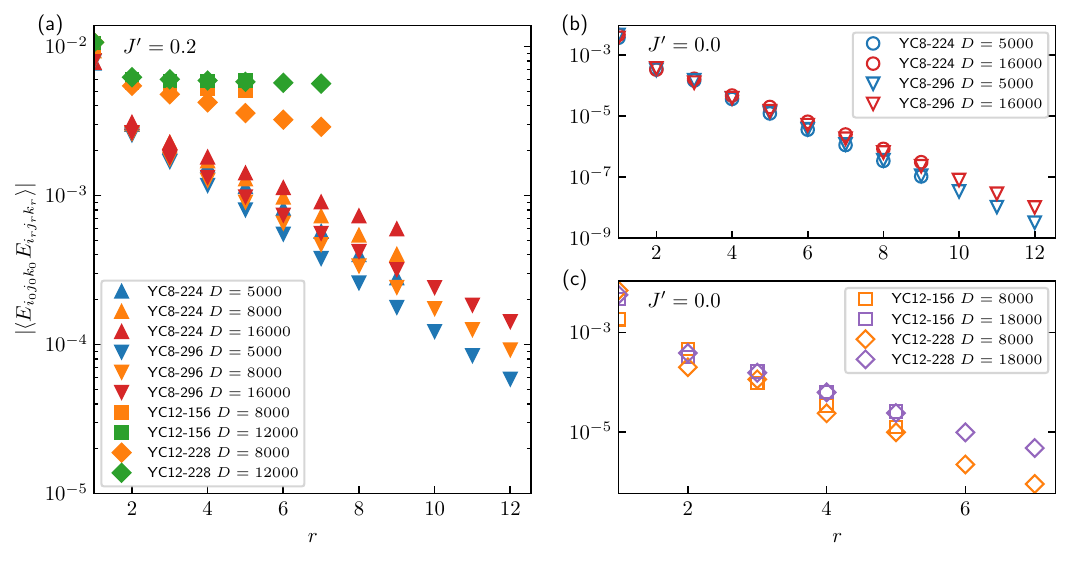}
\caption{\label{fig:cc_scaling}
	Chiral-chiral correlation function as a function of distance $r$ for (a) $J' = 0.2$ on YC8 and YC12 cylinders, (b) $J = 0$ on YC8 cylinders, and (c) $J' = 0$ on YC12 cylinders. Note that, for YC8 cases, we choose the reference point to $x=N_x/4$ and measure the correlator within the middle half of the cylinder to minimize the boundary effort, and for YC12 cases, we fix the reference point to $x = 2$. Real number wavefunction is adopted.
}
\end{figure}

We further show details of the chiral-chiral correlation function obtained in our DMRG calculations for both well-defined CSL state (at $J'=0.2$) and $J'=0$ state with varying system geometries and MPS bond dimensions in Supplementary Fig.~\ref{fig:cc_scaling}.

At $J' = 0.2$, on wider YC12 cylinders, the correlation function is flat when the bond dimension $D$ is larger than 12000, establishing the well convergency of our calculation and the existence of a long-range chiral order. On YC8 cylinders, while the ground state at $J' = 0.2$ has been accepted as a CSL \cite{gong2014emergent,gong2015global}, the chiral-chiral correlation function still present a decaying fashion, indicating the difficulty of faithful evaluation of a CSL phase in the KHAF by simply considering chiral-chiral correlation function. We emphasize that the data in Supplementary Fig.~\ref{fig:cc_scaling} are quantitatively consistent with previous works \cite{gong2014emergent,gong2015global}.

At $J' = 0$, for all the systems considered in Supplementary Figs.~\ref{fig:cc_scaling}(b-c), the correlation continuously enhances with an increasing bond dimension, implying the existence of a long-range chiral order. Notice that these data are also quantitatively the same as the data provided in Ref.~\cite{Depenbrock2012}. Moreover, because of a weaker and weaker chirality when approaching $J' = 0$ from a positive $J'$ (which implies a smaller and smaller energy gap), the required system size to observe a flat chiral-chiral correlation function is larger and larger, which makes it difficult to be accessible in DMRG calculations.

\subsection*{Supplementary Note 3: More details for the semion sector}
In this section, we provide more details on the characterization of the ground state in the semion sector. 

\begin{figure}[!ht]
\centering
\includegraphics[width = 0.8\linewidth]{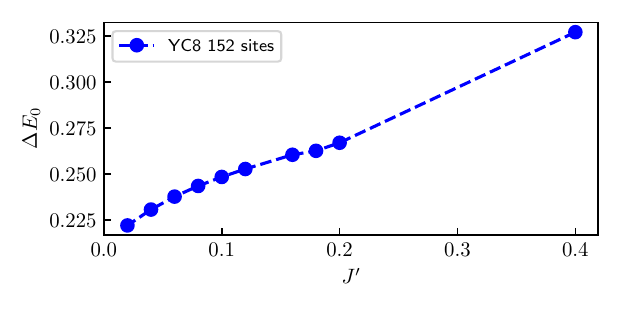}
\caption{\label{fig:csl_finite_size_gap}
	The finite energy gaps between $|\Psi_{\text{I}}\rangle$ and $|\Psi_{\text{S}}\rangle$ on a 152-site YC8 cylinder.}
\end{figure}

\textit{Finite-size gap between identity and semion sectors.---} Although the ground states in the identity and semion sectors, namely, $|\Psi_{\rm I}\rangle$ and $|\Psi_{\rm S}\rangle$, are degenerate in the thermodynamic limit, a finite energy gap between these two states exists for finite-size systems. In Supplementary Fig.~\ref{fig:csl_finite_size_gap}, we show that this finite-size gap decreases with $J'$.

\begin{figure}[!ht]
\centering
\includegraphics[width = 0.8\linewidth]{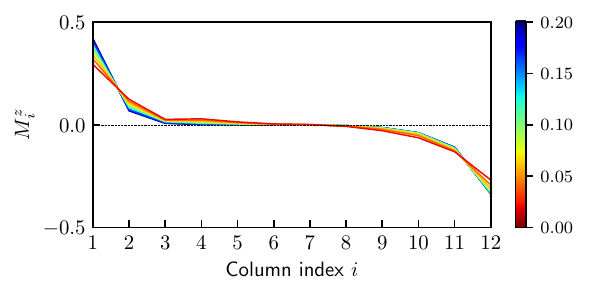}
\caption{\label{fig:kasyc12_8_semion_sz}
	The evolution of the column summation of the $M^z_i$ distribution of semion sector states in the 152-site YC8 cylinders from $J' = 0.2$ to $J' = 0.02$ with $D = 5000$. The coupling $J'$ for each curve can be read from the colorbar.}
\end{figure}

\begin{figure}
\centering
\includegraphics[width =0.8 \linewidth]{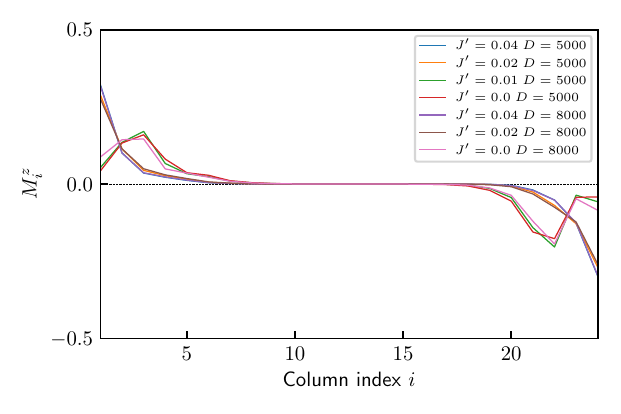}
\caption{\label{fig:bs_sz}
	The same as Supplementary Fig.~\ref{fig:kasyc12_8_semion_sz} but for doubled length 296-site YC8 cylinders with bond dimension up to $D = 8000$.}
\end{figure}

\textit{Localization length of the edge semions.---} In order to examine the edge semions,
we measure the total magnetization on the $i$-th column, dubbed $M^z_i$, with respect to the ground state in the semion sector.
The explicit definition of $M^z_i$ reads
\begin{equation}
M^z_i = \sum_{j}\langle{}\Psi_{\text{S}}|S^{z}_{i,j}|\Psi_{\text{S}}\rangle,
\end{equation}
where $i$ is the column index of the kagome lattice unit cell, and the summation $\sum_{j}$ runs over all the lattice sites belonging to the $i$-th column. In Supplementary Figs.~\ref{fig:kasyc12_8_semion_sz} and \ref{fig:bs_sz}, we show the $M^z_i$ profile as a function of $J'$ measured on 152-site and 296-site YC8 cylinders, respectively. Clearly, the localization lengths of two semions at the boundaries of the cylinder increase as $J'$ decreases. 
On the relatively short YC8 cylinders with $N=152$ (see Supplementary Fig.~\ref{fig:kasyc12_8_semion_sz}), the localization length of each edge semion might be as large as half of the cylinder length when $J'$ is sufficiently small, resulting in a breakdown of the adiabatic evolution of the semion sector around $J'=0.02$.
On YC8 cylinders with $N=296$ sites, i.e., the length of the cylinder has been doubled, the two semions localized at each boundary are separated further, and the semion sector therefore can 
be adiabatically evolved to the $J^\prime=0$ point. When $J'<0.02$, the edge semions start to penetrate into the bulk, and the highest peak of the corresponding $M^z_i$ does not locate exactly at the boundaries of cylinders. Note that the left (right) edge semion is more localized on the left (right) boundary of cylinders as the bond dimension increases from $D=5000$ to $D=8000$, indicating that the semion sector is robust against bond-dimension scaling at $J' = 0$.

\begin{figure}
\centering
\includegraphics[width = 0.8\linewidth]{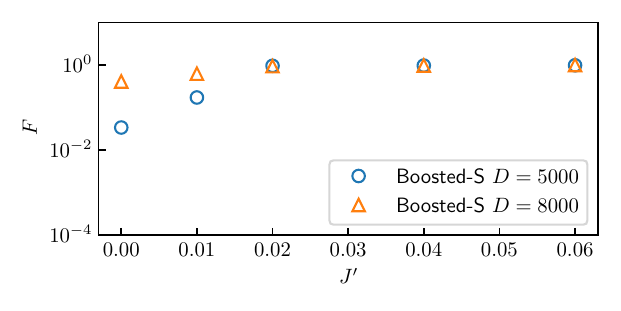}
\caption{\label{fig:ovlps_s_and_i}
	The neighboring wave function fidelity as a function of $J'$ in the semion sector. The data are obtained by Boosted-DMRG on 296-site YC8 cylinders with bond dimension up to $D = 8000$.
}
\end{figure}

\begin{figure}
\centering
\includegraphics[width = \linewidth]{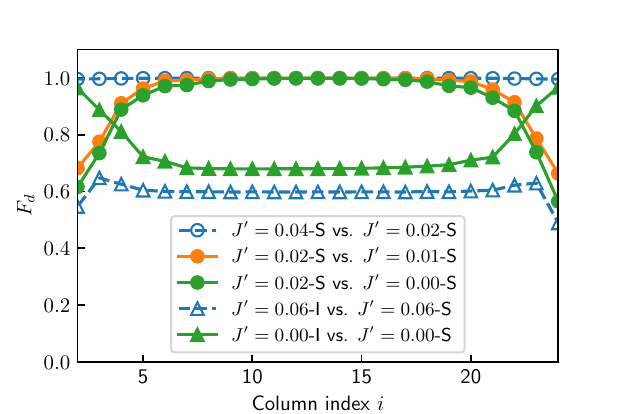}
\caption{\label{fig:dovlps_s_and_i}
The density-matrix fidelity [see definition in Eq.~\eqref{eq:Fd}] for each column $i$ on a 296-site YC8 cylinder. Here $J'=0.06$-I (-S) indicates the ground state obtained by Boosted-DMRG at $J'=0.06$ for the identity (semion) sector.}
\end{figure}

\textit{Fidelities in the semion sector.---} 
The neighboring wavefunction fidelity $F$ introduced in the main text is a direct metric for the adiabatic evolution of the semion sector. 
As shown in Supplementary Fig.~\ref{fig:ovlps_s_and_i}, $F$ is very close to 1 until $J'<0.02$ and there is a large jump of $F$ around $J'=0.01$. However, we emphasize that this jump of $F$ does not correspond to a phase transition at $J'=0.01$. By definition, $F$ is highly sensitive to the local information of two wave functions and below we will show that the large jump is caused by the slight deviations of semions from the boundaries of cylinders.

We find that $F$ at $J'=0$ increases significantly as the bond dimension increases from $D=5000$ to $D=8000$. As mentioned above, the edge semions are more localized with larger $D$, implying that the increase of $F$ results from the shift of semions to the boundary sides.

The absence of phase transition around $J'=0.02$ can also be revealed by studying the density-matrix fidelity $F_{d}$~\cite{McCulloch08},
\begin{equation}
F_{d}(\rho_{A}, \rho_{B}) = \mathrm{Tr}\sqrt{\rho_{A}^{1/2} \rho_{B} \rho_{A}^{1/2}}~,\label{eq:Fd}
\end{equation}
where $\rho_{A}$ and $\rho_{B}$, being two reduced density matrices for a column of the cylinder, correspond to two different states $|\Psi_{A}\rangle$ and $|\Psi_{B}\rangle$. This fidelity measures how close a specific part of two wave functions is, while the boundary effects are precluded as much as possible. $F_{d}$'s for the semion sector on each column are illustrated in Supplementary Fig.~\ref{fig:dovlps_s_and_i}. 
We find that $F_{d}$'s between the semion states at $J' = 0.02$ and $J' = 0.01$ as well as $J'=0$ (orange and green closed circles, respectively) are close to 1 in the middle of the cylinders, indicating the bulk parts of those states are quite close. However, these $F_{d}$'s around two boundaries are only about 0.6, due to the aforementioned slight shifts of edge semions. 
For comparison, we also show the $F_{d}$'s between two truly orthogonal states, e.g., the semion and identity sectors. The wavefunction fidelities (i.e., $F$'s) between those two states at the same $J'$ are round $10^{-10}$, and corresponding $F_{d}$'s are only round $0.65$ in the middle of cylinders. 



\subsection*{Supplementary Note 4: Performance of the Boosted-DMRG initialized with various parton ansatz}

\begin{figure}[!ht]
\centering
\includegraphics[width = 0.8\linewidth]{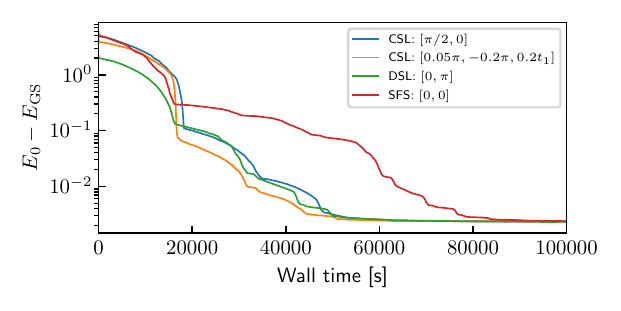}
\caption{\label{fig:b_cov_csls_dsl_sfs}
	Variational energy (minus a constant of $E_{\mathrm{GS}}$) versus the DMRG running wall time for different initial states in the $J' = 0$ calculations on 156-site YC8 cylinders with fixed $D = 5000$. Parameters employed in the parton Hamiltonian are shown in the legend. The $E_{\mathrm{GS}}$ is taken as the variational energy in the converged Random-DMRG calculation with $D = 8000$.}
\end{figure}

In addition to the $[\frac{\pi}{2}, 0]$ CSL state (employed in the main text as the initial state for the Boosted-DMRG calculations), we also prepare the Gutzwiller projected wave functions of $[0, \pi]$ Dirac spin liquid (DSL) state, $[0, 0]$ spinon Fermi surface (SFS) state, and an alternative CSL state with a $U(1)$ flux of $0.05\pi$ through each elementary triangle.
In the VMC calculation~\cite{Hu2015}, this alternative CSL, dubbed $[0.05\pi, -0.2\pi, 0.2t_{1}]$ here, was found to be the best variational ansatz for the established CSL phase in the $J-J'$ model, and ``$0.2t_{1}$'' stands for the additional 2nd NN hopping with strength $t_{2}=0.2$ in the parton Hamiltonian.

The above four Gutzwiller projected parton ansatz are employed to initialize  the Boosted-DMRG calculations at $J' = 0$,  in which all the simulation parameters such as bond dimension and maximal accepted Lanczos error have been chosen to be the same.
In Supplementary Fig.~\ref{fig:b_cov_csls_dsl_sfs}, we show the optimized ground-state energies as a function of the total wall time for the Boosted-DMRG calculations with different initial \protect{\it Ans\"atze}. At $t = 0$, the $[0, \pi]$ state has the lowest energy which is consistent with the previous VMC calculations. However, along with the DMRG sweep, the energy in calculations initialized by CSL states decays faster and offsets the higher initial energy drawback. Finally, the calculations initialized by the DSL state and two CSL states converge to \textit{the same} state at a similar time, implying that the $[0, \pi]$ state may not be the better \textit{Ansatz} for $J' = 0$ ground state although it has the best energy on the $J' = 0$ Hamiltonian.


\subsection*{Supplementary Note 5: Bond-bond correlation pattern in VBC phases}
\begin{figure}[!ht]
\centering
\includegraphics[width = 0.9\linewidth]{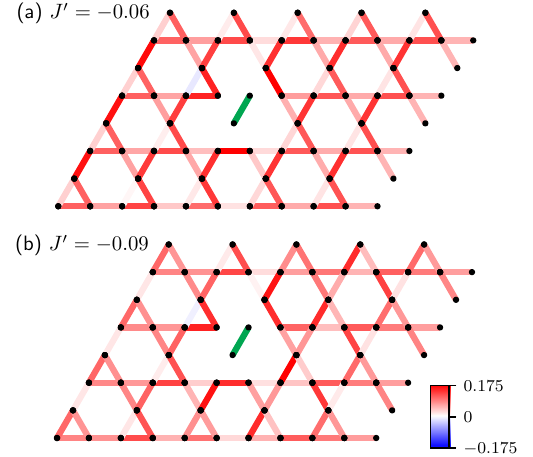}
\caption{\label{fig:bb2_vbcs} The bond-bond correlation $\langle(\mathbf{S}_i\cdot \mathbf{S}_j)(\mathbf{S}_m \cdot \mathbf{S}_n)\rangle$ patterns in (a) VBC II phase with $J^\prime=-0.06$ and (b) VBC I phase with $J^\prime=-0.09$. The reference bond ($i$-$j$) is indicated by green.}
\end{figure}

To investigate the VBC orders in the further detail, we calculate the bond-bond correlation functions in the VBC I and VBC II phases. The bond-bond correlation function is defined as $\langle(\mathbf{S}_i\cdot \mathbf{S}_j)(\mathbf{S}_m \cdot \mathbf{S}_n)\rangle$, where the NN bond $i$-$j$ is fixed to locate in the central of cylinders and $m$-$n$ can be other NN bonds away from $i$-$j$. 
As shown in Supplementary Fig.~\ref{fig:bb2_vbcs}(a), the pattern of $\langle(\mathbf{S}_i\cdot \mathbf{S}_j)(\mathbf{S}_m \cdot \mathbf{S}_n)\rangle$ for VBC II phase is consistent with the bond texture shown in Fig. 3(f) in the main text, implying that the VBC II phase on cylinders in principle preserves the translational symmetries in both directions. In the VBC I phase, the corresponding pattern shows that it breaks the translational symmetry in the $x$-direction by enlarging the unit cell. 

\newpage
\subsection*{Supplementary Note 6: System-size and bond-dimension scaling.}

\begin{figure}[!ht]
\centering
\includegraphics[width = 0.8\linewidth]{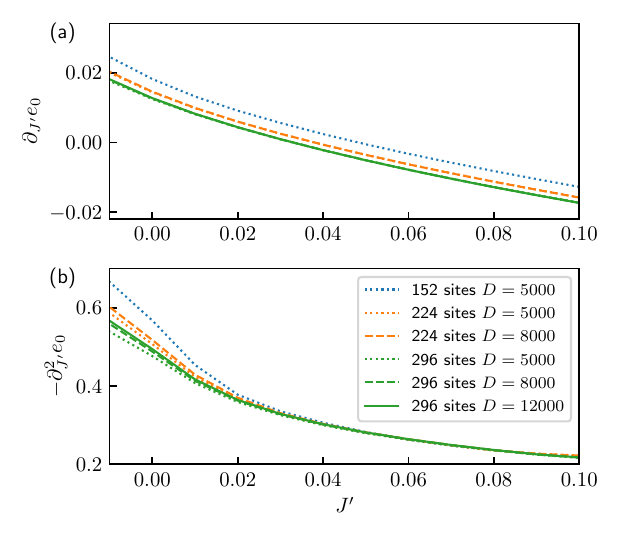}
\caption{\label{fig:e0_div2_scaling}
The scaling behavior of energy-density derivatives by changing the bond dimension $D$ and system size $N$. These results are obtained by Random-DMRG.}
\end{figure}

\begin{figure}[!ht]
\centering
\includegraphics[width = 0.8\linewidth]{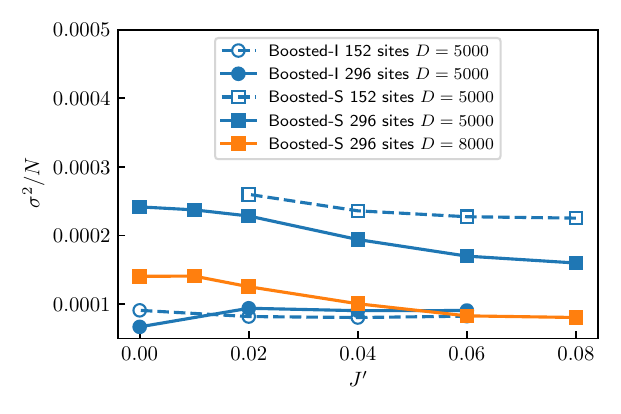}
\caption{\label{fig:sgms_vs_jps}
	The energy variance as a function of $J'$ for the states obtained by Boosted-DMRG on 152-site and 296-site YC8 cylinders up to bond dimension $D = 8000$.  Here Boosted-I (Boosted-S) stands for the ground state in the identity (semion) sector.}
\end{figure}

We perform more calculations by varying system size $N$ (up to $N=296$ sites) and bond dimensions $D$ (up to $D=12000$). The scaling behaviors of the first-order and second-order derivatives of the ground-state energy are shown in Supplementary Fig.~\ref{fig:e0_div2_scaling} for the most interesting regime ($0\leq J'<0.1$). We find that they are indeed smooth and do not show any singularity. It also shows that $\partial_{J'}^2 e_0$ ($e_0=E_0/N$ being the ground-state energy density) converges as $N$ and $D$ increase, indicating the robustness of our results against finite-size scaling. We also emphasize that Random-DMRG and Boosted-DMRG lead to the same result.

The accuracy of our Boosted-DMRG calculations can be further confirmed by the scaling behaviors of energy variances for the ground states in both the identity and semion sectors, see Supplementary Fig.~\ref{fig:sgms_vs_jps}. 
The energy variances for both sectors are of $10^{-4}$. It is straightforward to see that the energy variances for the semion sector significantly decrease as the system size and bond dimension increase, indicating that our DMRG simulations are safely converged and our results are robust.


\end{widetext}
\end{document}